
\newcount\fcount \fcount=0
\def\ref#1{\global\advance\fcount by 1 \global\xdef#1{\relax\the\fcount}}
\magnification=\magstep1

\raggedbottom
\footline={\hss\tenrm\folio\hss}

\tolerance = 30000

\baselineskip=5.00mm plus 0.1mm minus 0.1mm
\centerline {\bf Did Cosmic Rays Reionize the Intergalactic Medium?}
\vskip 2.5cm
\centerline {\bf Biman B. Nath and Peter L. Biermann}
\centerline{\it Max Planck Institut f\"ur Radioastronomie,
Auf dem H\"ugel 69, D-5300 Bonn 1, Germany}
\vskip 2.75cm
\centerline {\bf Abstract}
We investigate the role of cosmic rays from young galaxies in heating
and ionizing the intergalactic medium (IGM) at high redshift. Using
the IRAS observations at $60 \mu m$, we estimate the cosmic ray
luminosity density at the present epoch. We consider various forms
of luminosity evolution in redshift and calculate (a) the thresholds
corresponding to the upper limits of Gunn-Peterson optical depth,
(b) the Compton $y$ parameter for an IGM heated by cosmic rays and
compare with the upper limits from COBE measurements and (c) an
estimated limit from the integral of metal enrichment. We show that
certain models, with rather strong evolution and early formation of
galaxies, allow reionization of the IGM, consistent with all known
constraints.
\vskip 2.0cm
keywords -- cosmology : theory; galaxies : evolution, intergalactic medium.
\vskip 0.5cm
\centerline{To appear in MNRAS}
\vfill\eject
\noindent
{\bf 1. Introduction}
\medskip
The spectra of quasars provide us with
fascinating clues about the intergalactic medium (IGM) at high
redshift.
The well-known absence of Gunn-Peterson H Ly$\alpha$
absorption troughs in such spectra is the evidence that the
IGM has been highly ionized since, at least,
the epoch of the highest redshift quasars  ($z \sim 5$)
(Gunn \& Peterson 1965;
Steidel \&  Sargent 1987; Jenkins \& Ostriker 1991). On the other hand, in
the standard model
of big bang cosmology, the universe recombined at a redshift
of $\sim 1100$, leaving only a residual fractional ionization $\sim
10^{-4}$ (Peebles 1968).
To reconcile the theory and the observations,
astrophysicists over the years have suggested various models of reionization
of the IGM.

Quasars and young galaxies have been considered as candidate
sources for producing ionizing radiation needed to photoionize the IGM.
A metagalactic UV radiation has also been inferred from the ``proximity
effect'', the measured decrease in the number of Ly$\alpha$
clouds in the neighbourhood of a quasar due to its ionizing radiation.
It seems
that the uncertainties in our knowledge of quasars, galaxies at
high redshift and the Ly$\alpha$ clouds are too large to either confirm or
rule out
the scenario (Miralda-Escud\`e \& Ostriker 1990; Madau 1992;
but also see Shapiro \& Giroux 1987).
Apart from galactic radiation, heating of the
IGM by cosmological blastwaves (Ostriker \& Ikeuchi 1983) and decaying
massive neutrinos (Sciama 1990) have also been suggested as the possible
causes of ionization.

In this paper, we ask the question whether cosmic radiation from young
galaxies at high redshift played any role in reionizing
the IGM.  Star formation in the early universe almost certainly
produced cosmic ray particles through various acceleration
mechanisms. Unlike photons the cosmic ray particles
can ionize neutral atoms many times as they move in the IGM.
Besides, the gas can be collisionally ionized after being initially
heated by cosmic rays. The question is also
motivated from the point of view of energetics.
Miralda-Escud\`e \& Ostriker (1990), in their study of ionizing radiation
from young galaxies,
estimated that $\sim
10^{-3} \> M_{\odot} c^2$ of energy in ionizing photons is emitted
for each $1 M_{\odot}$ of metals produced. Absorptions inside and
outside the galaxy, and other losses, though, tend to attenuate the radiation
intensity to some extent. If we note that $0.1M_{\odot}$ of metals is
produced per $10^{51}$ ergs in supernovae shockwaves and
assume that
$\sim 0.1$ of the total shockwave energy is in the form of cosmic rays,
we get $ \sim 6
\times 10^{-4} \> M_{\odot} c^2$ per $1M_{\odot}$ of metals produced.
Naturally,only a fraction of this energy is finally available for heating the
IGM,
as in the case of photons, but the estimate indicates
that cosmic rays could be important as well.

It is interesting to note that Ginzburg \& Ozernoy (1965) had already
considered an IGM heated by cosmic rays from young galaxies, before the
Gunn-Peterson test was applied to the quasar spectra.
They used
cosmic ray energy densities of $10^{-15 \pm 1}$ erg cm$^{-3}$ and
speculated upon a highly ionized universe.
In the three decades since their work, our knowledge of the
universe at redshifts $z \la 5$ has increased dramatically and better
observational
constraints on the physical state of the IGM at
these redshifts have been obtained. In this paper, we consider the
problem of the cosmic ray heated IGM in the light of the new data. We
estimate the cosmic ray energy density from the IRAS observations of
the $60 \mu m$ luminosity function of galaxies, include
heating and cooling processes previously neglected. We present the
results in the form of threshold contours for various observational
limits in the plane of the IGM density and evolution parameters.

The organization of the paper is as follows. In section 2. we set up
the necessary equations, discuss the spectrum and the energy density of
cosmic rays.
In section 3, we calculate the heating and
ionization and then discuss various implications of an
IGM heated by cosmic rays in section 4.

\bigskip
\noindent
{\bf 2. Preliminaries}
\medskip
{\bf 2.1 Equations for ionization by cosmic rays}
\medskip
It is almost certain that young galaxies at high redshift produced
cosmic rays as the Milky Way does now. It is, however, an important point
as to how much of the energy in the form of cosmic rays leaked out
of the galaxies to the intergalactic medium, and how. The effect
of cosmic rays on the IGM depends crucially on the spectrum, the lower
energy cutoff and the total cosmic ray luminosity that finally emerges
from the galaxies.

The fraction of the total brightness in cosmic rays that
is put into the IGM and the lower cutoff of energy, depends
on the mode of leakage from the galaxies.
One possibility is that
of their being carried by galactic winds, in which case the particles will
lose energy due to adiabatic expansion ($p \sim r^{-1}$). This
mode of transportation lowers the cutoff energy as well as the
total energy.
We define $\beta_l$ as the lower cutoff in $\beta$ ($=v/c$)
of the particles after emerging from the galaxies.

Other important variables in the process of IGM heating
by cosmic rays are (a) $h$ (Hubble constantat the present era
is defined as $H_o=100 h$ km s$^{-1}$ Mpc$^{-1}$),
(b) the global $\Omega$ of the universe,
(c) $\Omega_{IGM}$, the density
of the IGM in the units of critical density
($\rho_c=3H^2/8 \pi G$),
(d) $z_s$, the epoch of galaxy formation,
and (e) the evolution of cosmic ray luminosity of galaxies.
$\Omega_{IGM}$ is certainly less than $\Omega_b$, the
density in baryons, which, according to current
estimates based on primordial nucleosynthesis (Kolb \& Turner 1990), is
equal to $(0.06 \pm 0.02) (h/0.5)^{-2}$. For simplicity, we consider an
$\Omega_{IGM}$ which is constant in time.

Consider the IGM in which cosmic rays with
luminosity $f_{cr} \> L_{cr} (z)$ (erg cm$^{-3}$ s$^{-1}$) is deposited by the
galaxies beginning at $z_s$. Here $f_{cr}$ is the dilution factor due to
the adiabatic energy loss of cosmic rays in the wind.
The particles have
a $p^{- \alpha}$ spectrum and we express their differential number
density by ${n_{cr} (\beta _i) \over E(\beta_l)}
\> d \beta _i$, where $\beta _i$
is the initial $\beta$ of a particle (i.e. before it interacts with
the IGM) and $E(\beta _l)=\int_{\beta_l}^1 E(\beta_i) n_{cr}
(\beta_i) d\beta_i$ in the denominator normalizes the spectrum.

The particles lose energy as they move in the IGM due to various
processes as we will soon explain in detail. The energy
of a particle, at any instant $z$, therefore depends on
(a) $z_e$ ($z<z_e <z_s$), the redshift at which it was emitted, (b)
its original energy, or, rather, in terms of velocity,
$\beta _i$ and (c) $z$, the current redshift. In the following sections, we
will inject
cosmic rays with certain initial energy density,
follow their evolution in time and
at each time step, integrate the effect of all the cosmic rays injected
prior to that instant. The particles
ionize the medium  with rate, $n_{HI} (z) \sigma \lbrack \beta (z)
\rbrack \beta (z)
c$. Here $n_{HI} (z)$ is the density of neutral atoms, $\sigma$ is the
ionization cross-section and $\beta \>(=v/c)$ denotes the
velocity of the particles.

The evolution of the energy density of IGM, $\epsilon (z)$ (erg/cm$^3$),
is governed by heating due to ionization ($\Gamma_{ion}$), heating
from the direct collision of cosmic ray particles with the free
electrons ($\Gamma_{cr}$) and cooling due to recombination, adiabatic
expansion of the universe and line emission of
hydrogen atoms ($L(HI)$). The ion-electron relaxation timescale is
$\sim 10 \> T^{3/2} n^{-1}$ sec, which is smaller than all other relevant
timescales.
We therefore use $\epsilon= {3nkT \over 2}$ as the definition of temperature
of the IGM.

The evolution of $f$, the
ionization fraction of the IGM is succinctly expressed by the
following set of equations, viz.,

$$\eqalignno{
n_{tot}(z) {df \over dt}&=\int\limits_z^{z_s} f_{cr} L_{cr}(z_e)
\Bigl( \int\limits_{\beta_l}^1 d \beta _i \> {n_{cr} (\beta _i)
\over E(\beta_l)}
\lbrace n_{HI}(z) \sigma \lbrack \beta (z,z_e,\beta _i) \rbrack
\beta (z,z_e,\beta _i)
c \rbrace \Bigr) dz_e &\cr
&+7.8 \times 10^{-11} n_e n_{HI} T^{1/2} \exp ({-13.6 {\rm eV} \over kT})
-\alpha_{rec} n_e n_{HI} ,&(1) \cr}
$$
$$ \eqalignno{
- {d \beta \over dt}&=1.94 \times 10^{-16} n_{HI}(z)
\> {(1- \beta ^2)^{3/2}
\over \beta} {2 \beta ^2 \over ((0.01)^3 +2 \beta ^3)}
( 1 + 0.0185 ln \beta H\lbrack \beta -0.01 \rbrack )
&\cr &+
3.27 \times 10^{-16} n_e(z)\>{(1- \beta ^2)^{3/2} \over \beta}
{\beta ^2 \over x_m^3+ \beta ^3}+
{\beta \over (1+z)} {dz \over dt}  ; &\cr
& x_m=0.0286(T/2 \times 10^6 K)^{1/2}, &(2)\cr}
$$
$$
{d \epsilon \over dt}= - \Bigl( {\epsilon \over t_{rec}} \Bigr)
- \Bigl( {5 \epsilon \over (1+z)} {dz \over dt} \Bigr)
-L(HI)+\Gamma_{cr}+\Gamma_{ion} \eqno(3)
$$

$$
{dz \over dt}= {1 \over H_o} {1 \over (1+z)^2 (1+\Omega z)^{1/2}}
\eqno(4)$$

The ionization cross-section for protons in equation
$(1)$ is given by (in the units of $\pi a_o^2$, where $a_o=5.3
\times 10^{-9}$ cm, is
the Bohr radius and $x$, the kinetic energy in keV)
$$\eqalignno{
\sigma &={1.393 \times 10^{-4} \over \beta ^2}
\Bigl( 6.2+ Log {\beta ^2 \over (1- \beta ^2)}
-0.43 \beta ^2 \Bigr), \quad \beta  \ge 0.026 &\cr
&=3.455-0.0386 x+0.01347 x^2- 2.463 \times 10^{-3} x^3+ 1.75 \times
10^{-4} x^4, &\cr \qquad &\qquad 0.0115  < \beta < 0.026 &(5)\cr}
$$
The expression in the first line of equation $(5)$ is given by
Spitzer \& Tomasko (1968) and that in the second line
is an analytical fit for the data given by Fite et.
al (1960).
Following Spitzer \& Tomasko, we multiply the cross-section by
${5 \over 3}$ to allow for ionizations by the secondary electrons,
and by an additional factor of $1.89$ for ionizations by heavy ions.
The factor of ${5 \over 3}$, to be precise, is not a constant and varies
with the fractional ionization ($1.67$ for $f \sim 0$ and $\sim 1.12$
for $f \sim 0.3$ (Spitzer and Scott 1969)), but the discrepancy is
expected to be of the order of unity.

The second term on the right hand side of equation $(1)$ is due to collisional
ionization in a gas with temperature $T$.
The last term denotes recombination with $\alpha_{rec}=
2 \times 10^{-11} T^{-1/2}$ cm$^3$ s$^{-1}$.

The first term in equation $(2)$ represents the loss of energy of cosmic rays
due to ionization ($H$ is the Heaviside function) while the second term
denotes the loss of energy
due to Coulomb interactions with free electrons
(Mannheim and Schlickeiser 1993). The last term
is due to the expansion of the universe.

The heating due to ionization, $\Gamma_{ion}$, depends on the mean
energy of the electrons and the probability of energy loss in inelastic
collisions. For small values of $f$, $\Gamma_{ion}=5 \times 10^-{12} I$
erg cm$^3$ s$^{-1}$, where $I$ (cm$^{-3}$ s$^{-1}$) is the rate of ionization,
and for $f>0.1$,
$\Gamma_{ion}$ is an order of magnitude higher. However, $\Gamma_{cr}$,
the heating due to direct collisions of cosmic rays with free electrons,
dominates all heating processes for $f \ga 0.1$.
We use $\Gamma_{cr}=- \int {dE(\beta) \over dt} n_{cr} (\beta) d \beta$, where
${dE(\beta) \over dt}$ is the Coulomb loss for a cosmic ray particle
with velocity $c \beta$ from the second term on the right hand side
of the equation $(2)$.
$L(HI)$ is the rate of line cooling
by hydrogen atoms at $937.8, 949.7, 972.5 \AA$ (Gaetz and Salpeter
1983).

\bigskip
{\bf 2.2 Cosmic Rays -- energy density, spectrum and leakage}
\medskip

The luminosity function of galaxies at $60 \mu m$ provides an estimate
for $L_{cr}(z=0)$. The luminosity in far infrared wavelengths
can be scaled to that in cosmic rays from the observations of M82.
Kronberg et al (1985) estimated a supernova rate of $\sim 1$ per $3$
years in M82 with an uncertainty of a factor of three. Assuming
that the efficiency of producing cosmic rays is about $10\%$ and
the energy input per supernova is $ 10^{51}$ ergs, we obtain the cosmic
ray luminosity of $10^{42}$ ergs s$^{-1}$. Comparing this with the $60 \mu m$
luminosity of M82 of $4 \times 10^{10} L_{\odot}$ (Rieke et al 1980), we get
a ratio of $0.007$ between the cosmic ray and far infrared luminosity. We
will use a ratio of $0.01$ and discuss the effect of the uncertainties
in section 3.5.

Lawrence et. al.
(1986) fitted their IRAS data with the luminosity function,
$$
\phi (L) =CL^{-1} (1+{L \over L_\ast  b})^{-b}, \eqno(6)
$$
where $\phi (L)$ is defined such that $\phi (L) dlog_{10}L$ is the
number of sources per Mpc$^{-3}$ in the luminosity range $log_{10} L,
log_{10} L+dlog{10}L$. The unit of $L$ is
$L_{\odot}=3.9 \times 10^{33}$ erg s$^{-1}$. They obtained their best
fit with $b=2.4$, $log L_\ast=11.3$, $logC=7.12$. Integration
with their observed lower cutoff at $log_{10}L \sim 7.5$ yields a
luminosity density of $1.6 \times 10^{-32} h$ erg cm$^{-3}$ s$^{-1}$. The
scaling discussed above then gives a cosmic ray luminosity density of
$1.6 \times 10^{-34} h$ erg/s/cc at $z=0$. This is the value we adopt
for $L_{cr} (z=0)$ in equation $(1)$.

The spectrum of cosmic rays outside the galaxy is the same as the
source spectrum in the simple leaky box argument. While there are obvious
problems with such an argument (see Biermann 1993), there are no better
alternatives at present. Biermann (1993) (and other papers in the series) has
discussed the acceleration of cosmic rays and the predictions have been
verified with
airshower data. The basic idea discussed in these papers is that
cosmic rays (a)  up to about 10 TeV particle energy (for hydrogen) are
dominated by normal supernova explosions in the interstellar medium,
(b) from 10 TeV to near EeV particle energies are dominated by
supernova explosions into stellar winds, and (c) beyond EeV
particle energies are dominated by radio galaxies.
For the low energy cosmic rays this means that their spectrum is
$ \sim E^{-2.4}$ at injection, and again outside the galaxy.
Note that this injection spectrum is the relativistic approximation. The
theory of shock acceleration (e.g., Drury 1983), however,
tells us that the injection spectrum is actually a power law in momentum
across the transrelativistic region, i.e., $p^{-2.4}$.

For the lower energy cutoff in the spectrum there are two
arguments. First, one needs a lower cutoff in the range of $30-100$ MeV
bundances in intersteller clouds (for example, Black et al 1990)
repeating the arguments of Spitzer \& Tomasko (1968)
(Jokipii \& Biermann, in preparation).
Second, the production of Be and other elements from spallation by cosmic
rays indicates an energy cutoff in the same range (Gilmore et al 1992).

We thus adopt a cosmic ray spectrum of $\sim p^{-2.4}$ and a low kinetic
energy cutoff of $30$ MeV ($\beta=10^{-0.6}$).
We will discuss the consequences of the
uncertainties in these values in section 3.5.

Let us here note that in the case of galactic winds carrying the
cosmic ray particles, the fraction $f_{cr}$ of the total energy
that is put into the IGM, is a function of the ratio
between the lower cutoff in particle momentum inside and outside
the galaxy. With $p \sim r^{-1}$ ( where $r$ denotes length scale)
the lower cutoff and the flux ($n(p) \beta c dp$) decrease
as the particles lose energy adiabatically
expanding into the IGM.
The fraction $f_{cr}$ is simply the ratio of the energy contents of these two
particle spectra, inside and outside the galaxy.
After fixing the lower cutoff inside the galaxy,
$f_{cr}$ therefore becomes a function of $\beta_l$.

In a galactic wind, adiabatic loss of the cosmic rays dominates over
their ionization loss. The energy of the non-relativistic particles
scale as $E \sim p^2 \sim (r/r_o)^{-2}$, where $r_o$ is some fiducial
length scale. The density of particles is given by $n \sim n_o (r/r_o)^{-3}$
(from the conservation of number of particles in a spherically symmetric
wind). The number of times a cosmic ray particle has ionizing
collisions is $\sim \int_{r_o} ^{r_f} n \sigma dr$ which, for a final
radius $r_f \gg r_o$, is $\sim 0.5 n_o \sigma r_o$. A value of $\sigma
\sim 10^{-18}$ cm$^2$, $r_o=1$ kpc and $n_o=1$ cm$^{-3}$ should give
us an upper limit on the ionization loss. With $\sim 50$ eV lost per
ionization, this gives
$\sim 25$ keV, which is negligible for a $\sim$ few MeV particle
compared to the adiabatic loss. The low energy cosmic rays therefore
are carried by the galactic winds to the IGM with
energy lost mainly by the adiabatic expansion and we neglect the
ionization loss inside the galaxy and in the wind.

\bigskip
\noindent
{\bf 3. Results}
\medskip
{\bf 3.1 Approximate considerations:}
\medskip

Let us first try to estimate the heating by cosmic rays before solving
the equations exactly.
Following Ginzburg and Ozernoy (1965),
for cosmic ray energy densities of $w_{cr}$ (eV/cm$^3$) with all the protons
having energy $E_{cr}$,
the heating due to cosmic rays can be approximately written as,
$$
\Gamma \approx 10^{-11} n w_{cr}\Bigl ({E_{cr} \over 1 {\rm MeV}}
\bigr ) ^{-3/2}
\> {\rm {eV \over cm^3 \> sec}}.
\eqno(7)
$$
Here $n$ is the particle density. As we have seen in the previous
section, the local energy density in cosmic rays is around a hundredth
of that in $60 \mu m$, i.e., $w_{cr} \sim 10^{-5}$ eV cm$^{-3}$;
The dilution factor $f_{cr}
\sim 0.1$ for a decrease in the lower energy cutoff to
$\sim 1 $ MeV from an initial cut off of $\sim 30 MeV$ inside the
galaxy. With no luminosity evolution, a cosmic ray energy density
of $10^{-6}$ eV cm$^{-3}$ seems reasonable to be used in $(7)$.
Therefore, over a Hubble time ($ t_H \sim 4 \times
10^{17}$ sec for $h=0.5, \Omega=1$), the cosmic rays put $\sim 4$
eV per particle in the universe. However,
the approximation of using
all the protons at the lower energy end overestimates $\Gamma$
by an order of magnitude, as we will show in section 3.2, and the
energy per particle in this case is about $0.5$ eV.

Without any evolution in luminosities, the cosmic ray energy density thus
is not enough to heat up the IGM to high temperatures to ionize collisionally.
However, the above estimate shows that if $w_{cr}$ is larger
by a factor of $\sim 50$ due to moderate
evolution, the IGM can be heated up to $T \ga 10^5$ K. Collisional
ionization then will rapidly deplete the neutral atoms of hydrogen.

\bigskip
{\bf 3.2 Heating and Ionization of the IGM and the Gunn-Peterson Test:}
\medskip
The Gunn-Peterson optical depth due to neutal hydrogen at $z$ can be written
as,
$$
\tau_{GP}=4.6 \times 10^5 \Omega_{IGM} h (1-f) (1+z)^2 (1+\Omega z)^{1/2},
\eqno(8)
$$
where $(1-f)$ is the neutral fraction. It is evident that the optical depth
rises steeply with $z$ and the test becomes most sensitive at high
redshifts. The highest redshift at which $\tau_{GP}$
has been measured is at $z=4.2$ and the upper limit is
$\tau_{GP} < 0.14$ (for $h=0.5$) (Jenkins and Ostriker 1991).
Webb et. al's (1992) measurement of $\tau_{GP} = 0.04$ at $z=4$ is
dependent on the assumptions on the spectrum of Ly$\alpha$ clouds which are
yet to be confirmed and we use the former upper limit of $\tau_{GP}$.

We consider three forms of evolution and calculate the thresholds
for the above limit of $\tau_{GP}$ at $z=4.2$.

{\it Case I:} Here, we assume the luminosity function to have a
single power law, i.e.,
$$
L_{cr}(z)=L_{cr}(z=0)
(1+z)^{3+m}\qquad z \le z_s. \eqno(9)
$$
The threshold contours in the
$(\Omega_{IGM} - m)$ plane
are shown in  fig. 1(a) for $ 1+z_s=8,10$ and
$\Omega=0.1,1.0$.

{\it Case II:} In this case we assume the evolution to have a
``broken''
power law: with an index $m$
till a certain redshift $z_c$ when the evolution is ``switched off'',
and
the galaxies simply comove beyond $z_c$. That is,
$$\eqalignno{
L_{cr}(z)&=L_{cr}(z=0)
(1+z)^{3+m} \qquad z \le z_c &\cr &=L_{cr}(z=z_c) (1+z)^3
\qquad z_c < z  \le z_s. &(10)\cr}
$$
Recent observations
of high redshift quasars have indeed found their evolution to have a
``broken'' power
law luminosity evolution, with $z_c \sim 2-3$ and $m \sim 3.5$ (Boyle
1991).
Contours are shown in the $(\Omega_{IGM} - z_c)$ plane
for $1+z_s=10,8, m=4$ and $\Omega=0.1,1.0$ in  fig. 1(b)

{\it Case III:} Following Miralda-Escud\`e and Ostriker (1990), we consider
the case where galaxy formation rate per unit redshift is a Gaussian,
$\phi(z) dz \propto \exp \lbrack - {1 \over 2} ({z-z_f \over w})^2
\rbrack dz$. Conceivably, $L_{cr} (z)$ could also be a Gaussian of the
above form, i.e.,
$$\eqalignno{
L_{cr} (z) &\propto ({\rm Constant}) \exp\lbrack - {1 \over 2} ({z-z_f \over
w})^2 \rbrack \> (1+z)^3  &\cr
&= L_{cr} (z=0) \exp \lbrack - {1 \over 2 w^2} ( (z-z_f)^2 -z_f^2)
\rbrack \> (1+z)^3. &(11)\cr}
$$
Contours in the $(\Omega_{IGM}-w)$ plane
are shown in fig. 1(c) for
$1+z_s=1+z_f=10, 8$ and $\Omega=0.1, 1.0$.

It is evident from fig. 1 (a, b,c) that, in general, a lower value of $\Omega$
corresponds to contours with small evolution. This is
because in an older universe (i.e., lower $\Omega$) the cosmic
rays have more time to ionize and heat the IGM for the same range of redshift.
The curves also show that stronger galactic winds can heat and ionize
the IGM more easily.

As was pointed out in the last section, most of the heating occurs
after the cosmic rays ionize the IGM to $f \ga 0.1$. Heating by $\Gamma_
{cr}$ and collisional ionization then both act to raise $f$. At
temperatures $\sim 10^4$ K, line cooling of neutral hydrogen is
important but its effect diminishes with increasing $f$. The
recombination timescale ($\sim 5.0 \times 10^{17} (1+z)^{-3} h^{-2}
\Omega_{IGM}^{-1} (T/10^4)^{1/2}$ sec) is large compared to other
cooling
timescales and is therefore less important. Cooling
due to the expansion of the universe becomes important only at lower
redshifts. To illustrate the effects of various heating and cooling
mechanisms, we plot in fig. 2.,
the temperature of the IGM, various
cooling and heating terms as functions of redshift, for the case of
$L_{cr} \propto (1+z)^{m+3}, m=3$, $1+z_s=10, \Omega=0.1,
\Omega_{IGM}=0.01, h=0.5$.

It is useful to calculate the the fraction of the total energy
in cosmic rays that is lost to the IGM. For evolutions of
the type Case I above, we find that the
fraction depends on the lower cutoff $\beta_l$ and  $ \Omega, \Omega_{IGM}$
and it is fairly insensitive to the initial redshift $z_s$, the
evolution index $m$, the cosmic ray spectrum index $\alpha$.
{}From equation $(3.1)$, it is easy to see that the fraction $ {\Gamma_{cr}
\over
w_{cr}} t_H \propto \Omega_{IGM}$ and is bigger for an older universe
(i.e., smaller $\Omega$).
The exact dependence on $\beta_l$ and $\Omega_{IGM}$ is plotted in
fig. 3.
Equation ($7$) predicts, for example, for $\Omega_{IGM}=0.01, \Omega=1.,
h=0.5, \beta_l=10^{-1.3}$, that a fraction $\sim 0.1$ of the cosmic
ray energy density to be lost in moving through the IGM. Considering
the fact that for $\beta_l=10^{-1.3}, \> f_{cr} \sim 0.1$, the plot shows
that the approximate expression in equation ($7$) overestimates the energy
loss by an order of magnitude.

Theories of galaxy formation have not yet acquired the finesse to be
able to predict the form of evolution of the luminosity function.
Recent IRAS observations have
shown evidences (for example, Lonsdale, C. et.al 1990) for the evolution
index being $m=3-4$ with a cutoff at $z_s \sim 3$. The fit to the data
is fairly insensitive to the cutoff $z_s$ since mostly galaxies at $z<1$
contribute to the data.
As the curves of fig. 1
show, the reionization of the IGM by cosmic rays needs
the evolution index $m$ to be as large as this but with an
epoch of galaxy formation much earlier.
However, due to the
uncertainty in interpreting these data, we will not use them
as observational constraints. We will calculate the effects of
luminosity evolution on the IGM and compare with the constraints from
COBE
and abundances heavy elements in the universe in the next
sections.

\medskip
{\bf 3.3 COBE limit} \medskip

Recent measurement of the Compton $y$ parameter of the microwave
background radiation with COBE has put severe constraints on the history
of a hot IGM. An upper limit of
$$
y = \int {n_e k T \over m_e c^2} \sigma_{Th} c \> dt \> < 2.5 \times
10^{-5} \eqno(12)
$$
has been reported
(Mather et al 1993). A high cosmic ray
energy density could heat up the IGM to temperatures which are
ruled out by such a limit. We show in fig. 1 (a,b,c).
the curves corresponding to the above limit on $y$. The regions
bounded by these curves and the threshold curves for the upper
limit of $\tau_{GP}$ are the allowable regions for an IGM
heated by cosmic rays.

The curves show that, for Case I, with the single power law
evolution, the Gunn-Peterson and the COBE limit exclude an
hot and collisionally ionized IGM with $\Omega \ga 0.1$. For the
``broken'' power law case, however, the limits are not stringent
for $m=4$, the case we have considered. For higher values of $m$,
both Gunn-Peterson and COBE limit lines will shift towards
lower values of $1+z_c$. With a Gaussian form of evolution, COBE limit
curves, again, approach the $\tau_{GP}$ curves at high $\Omega_{IGM}$.

We found that for the points on the $\tau_{GP}$ threshold contours
for $\Omega_{IGM} \la 0.05$, the
values of the Compton $y$ parameter are about a hundredth of the
current upper limits.

\medskip
{\bf 3.4 He II Gunn-Peterson test:} \medskip

The optical depth due to singly ionized helium atoms in the IGM,
if observed in the near future, can put interesting constraints
on the physical state of the IGM and its history. The optical
depth for the Ly$\alpha$ line of HeII ($304 \AA$) is given by
$$
\tau_{HeII}=7.7 \times 10^2 {n_{HeII} \over n_{He}}
\Bigl ({\Omega_{IGM} \over 0.01} \Bigr ) ({h \over 0.5}) (1+z)^2
(1+ \Omega z)^{-1/2}, \eqno(13)
$$
assuming a $25 \%$ helium abundance. In ionization equilibrium
at $T \sim 10^{5.5}$ K, the fractional abundance of HeII is $\sim
10^{-3.3}$. Therefore, for an IGM at such a temperature at $z$,
the optical depth will be
$$
\tau_{HeII} \sim 3.9 \times 10^{-2} \Bigl ({\Omega_{IGM}
\over 0.01} \Bigr ) ({h \over
0.5}) (1+z)^2 (1+ \Omega z)^{-1/2}. \eqno(14)
$$
It is much smaller than that in most of the models considered by Miralda-
Escud\`e and Ostriker (1990) for a photoionized IGM.
The He II Gunn-Peterson test, however, is feasible only for redshifts more
than two for the wavelength of the photons need to be long enough
to avoid absorption in our Galaxy.

\medskip
{\bf 3.5 Uncertainties}
\medskip

It is easy to see the effects of using different values of the parameters that
we have used. The Gunn-Peterson limit curves in fig. 1 show the balance
between the effect of collisional ionization ($\propto n^2 \sim
\Omega_{IGM}^2
h^4$) and the optical depth ($\propto \Omega_{IGM} h$). The
curves, therefore, scale as $\Omega_{IGM} h^3$. In other words, for a change in
$h$ by a factor $a$, points in the curves at a certain $\Omega_{IGM}$ would
shift to $a^3 \Omega_{IGM}$. The COBE limit curves scale as $\Omega_{IGM} h$
since $y \propto n t_H$. The allowable regions between the curve, thus,
become narrower with increasing $h$ and make IGMs with higher
$\Omega_{IGM}$ more difficult
to reconcile with both $\tau_{GP}$ and $y$ limits.

The calibration of cosmic ray
luminosity from the observations of M82 has an uncertainty of
about a factor of three. As the effect of collisional ionization is
proportional to $n^2 w_{cr}$ the curves corresponding to $\tau_{GP}$
and $y$
scale as $a^{1/2} \Omega_{IGM}$ where $a$ is the factor of
uncertainty in
the luminosity. We have checked that numerical calculations
validate these qualitative arguments. The effect of changes in the
spectral index $\alpha$ for the cosmic rays by an amount $\pm 0.1$ is
small enough to be neglected. Also, a change by a factor of $\sim 3$
in the lower energy
cutoff corresponds to a change by $\sim 10^{0.23}$ in $\beta_l$ and
the change is expected to be of the same the order as between
the curves of different $\beta _l$.

\bigskip {\bf 4. Discussions} \medskip

As we noted in section $1$, the amount of heavy element enrichment
in the universe can be associated with the cosmic ray energy density. We
have already seen that $1\> M_{\odot} c^2$ ergs of energy
corresponds to the production of $1 M_{\odot}$ of metals produced in supernova
explosions. If we take a mean metal density of $2 \times 10^{-32}$ g cm$^{-3}$
(corresponding to $\Omega_b=0.1$ and mean metallicity equal to $0.02$)
we get an upper limit on the cosmic ray energy density of $10^{-14}$
ergs cm$^{-3}$. To show how this limit constrains the scenario of cosmic
ray heating of the IGM, we have drawn corresponding vertical lines in the
plots of fig. 1 (allowable regions are to the left of the plotted lines).
One must bear in mind that this constraint is at best a crude
one, depending on the assumptions of the mean metallicity, $\Omega_b$
and the fraction of supernova explosion energy going into cosmic
rays. If the estimates are not very far from reality, then they
give stronger limits than the COBE measurements.
Note that, since in this model one needs strong galactic winds to carry
the cosmic rays out to the
IGM, one expects a nontrivial enrichment of the IGM as well.

Another possible constraint is that the high energy tail of the intergalactic
cosmic rays originating in galaxies should not dominate over the galactic
cosmic rays since
heavy elements have been found to be abundant in that regime
(Stanev et. al 1993).

In the above calculation, we have considered an isotropic distribution
of cosmic rays and a homogeneous IGM. It is interesting to speculate upon
the effects of clumpiness in the cosmic rays as well as in the IGM.
The galactic winds carrying the cosmic rays will have to emerge out
of the clumps, which have higher densities, to produce any noticeable
heating and ionization. Cosmic rays could achieve this more easily than
the photons in a photoionized IGM.

Inhomogeneities in the IGM may lead to formations of pockets of gas
with high neutral fraction
(at high density pockets the gas would cool faster) and may show local
Gunn-Peterson absorption troughs. The inhomogeneities would also impart
anisotropies (${\Delta T \over T}$) in the microwave background
radiation of the order $\sim 2y$ (i.e., $\sim 5 \times 10^{-7}$ for
points on the threshold curves for $\tau_{GP}$) in the angular
scale that is appropriate for the clumps.

We have also treated the IGM density as a constant in time.
The result that ionization of the IGM is easier for very low IGM density
also features in the photoionization models. However, whether such
a low IGM density at high redshifts can be reconciled with galaxy
formation is an important question (Shapiro et al 1991).
Our results pertain only to the IGM density at redshifts $z \ga 4$, before
the era now accessible through the Gunn-Peterson test.
Future works on structure formation and evolution of galaxies should
shed more light on the problem and constrain the models better.

\bigskip
{\bf Conclusion}
\medskip
We have considered the heating and ionization of IGM by cosmic rays
from young galaxies. Using the IRAS luminosity function of galaxies
at $60 \mu m$ we estimated the local cosmic ray energy density to
be $ \sim f_{cr} 10^{-5}$ eV cm$^{-3}$, where $f_{cr}$ is the dilution
factor depending on the energy loss in escaping the galaxies into the IGM.
We have calculated the effect of low energy cosmic rays,
carried outside the galaxy by winds, for different models of
galactic evolution. We found that if the energy density were larger
(for example, by a factor of $\sim 50$, for $\Omega_{IGM}=0.01, \Omega=0.1,
h=0.5$) at high redshifts due to galactic evolution, heating and
ionization of the IGM by cosmic rays would be important.

The observations that constrain such a scenario are Gunn-Peterson
tests (for neutral hydrogen, and HeII, in the near future), Compton
$y$ parameter from COBE measurements and heavy element enrichment
in the universe. We found that COBE and Gunn-Peterson limits
provide the strongest constraints and we have shown that there
are various models of luminosity evolution of galaxies for which
both these limits can be satisfied. In such cases, the IGM is
partially ionized and heated to temperatures in excess of $10^5$ K which
then collisionally ionizes to the Gunn-Peterson limit.

We have calculated the Gunn-Peterson optical depth for HeII in
a hot IGM, which may provide, in the future, a test to distinguish
between a photoionized and a collisionally ionized IGM.

\bigskip
\bigskip
{\bf Acknowledgments}
\medskip
We thank Drs. V. Berezinsky, E. Kreysa, P. Shapiro and J. Silk for
illuminating discussions. BN wishes to thank Max Planck Society
for a postdoctoral fellowship. High energy astrophysics with
PLB is supported by the DFG (Bi 191/6,7,9), the BMFT (DARA FKZ
50 OR 9202) and a NATO travel grant (CRG 910072).
\vfill\eject
{\bf References}

\parskip=0pt
\parindent=0pt
\medskip

{\obeylines

1. Biermann, P. L. 1993  A\&A, in press.
2. Black, J. H. {\it et. al} 1990. ApJ, 358, 459.
3. Boyle, B. J. 1991. in The Space Distribution of Quasars,
D. Crampton (ed.), (San Francisco: Astron. Soc. of the Pacific), 191.
4. Drury, L.O'C 1983. Rep. Prog. in Phys., 46, 973.
5. Fite, W. L. {\it et. al} 1960. Phys. Rev., 119, 663.
6. Gaetz, T, Salpeter, E. 1983. ApJS, 52, 155.
7. Gilmore, G. {\it et. al} 1992. Nature, 357, 379.
8. Ginzburg, V. L., Ozernoy, L. M. 1965. Astron. Zh., 42,
943; Engl. transl. 1966., Sov. Astr. A. J., 9, 726.
9. Gunn, J. E., Peterson, B. A. 1965. ApJ, 142, 1633.
10. Jenkins, E. B., Ostriker, J. P. 1991. ApJ, 376, 33.
11. Kolb, E. W., Turner, M. S. 1990. The Early Universe (Addison-Wesley).
12. Kronberg, P. P., Biermann, P. L., Schwab, F. R. 1985 ApJ, 291, 693.
13. Lawrence, A. {\it et. al} 1986. MNRAS, 219, 687.
14. Lonsdale, C. {\it et. al} 1990. ApJ, 358, 60.
15. Madau, P. 1992. ApJ, 389, L1.
16. Mannheim, K., Schlickeiser, R. 1993. preprint.
17. Mather, J. C. {\it et. al} 1993. COBE preprint
18. Miralda-Escud\`e, J., Ostriker, J. P. 1990. ApJ, 350, 1.
19. Ostriker, J. P., Ikeuchi, S. 1983. ApJ, 268, L63.
20. Rieke, G. H. {\it et. al} 1980. ApJ, 238, 24.
21. Peebles, J. P. E. 1968. ApJ, 153, 1.
22. Sciama, D. W. 1990. MNRAS, 246, 191.
23. Shapiro, P. R., Giroux, M. L. 1987. ApJ, 332, 157.
24. Shapiro, P. R., Giroux, M. L., Babul, A. 1991. in After The
First Three Minutes, AIP Proceedings no. 222 (New York: AIP), 347.
25. Spitzer, L., Tomasko, M. G. 1968. ApJ, 152, 971.
26. Spitzer, L., Scott, E. H. 1969, ApJ, 158, 161.
27. Stanev, T., Biermann, P. L., Gaisser, T. K. 1993, A\&A, in press.
28. Steidel, C. C., Sargent, W. L. W. 1987. ApJ, 318, L11.
29. Webb, J. K. {\it et. al} 1992. MNRAS, 255, 319.
}

\vfill\eject
{\bf Figure Captions}
\medskip
Figure 1 (a): Thresholds contours for $\tau_{GP} < 0.15$ at $z=4.2$ and
those corresponding to $y(z=0) < 2.5 \times 10^{-5}$ are
plotted for the Case I evolution. The solid, dotted and dashed
curves are for $\beta_l=10^{-2}, 10^{-1.5}, 10^{-1.3}$ respectively.
(With an initial lower cutoff of $30 $ MeV, these curves correspond to
adiabatic loss of energy in winds by factors of $25, 7.7, 5.0$ respectively.)
The set of curves on the left and right correspond to the $\tau_{GP}$
and $y$ limits respectively and allowable regions are below the curves.
Case I, with a single power law evolution is considered here
for $\Omega=1, 0.1$ and $1+z_s=10, 8$.

The vertical dot and dashed lines correspond to the constraints from metal
enrichment. Allowable regions are to the left of these lines. The
horizontal long dashed lines are the limits on $\Omega_b$ ($0.006 \pm 0.02$)
from primordial nucleosynthesis ($\Omega_b > \Omega_{IGM}$).

\medskip
Figure 1(b): Contours for Case II,
with a broken power are shown for $\Omega=1, 0.1$
and $1+z_s
=10, 8$ and $m=4$. The curves are in the parameter space of
$\Omega_{IGM}--1+z_c$.

\medskip
Figure 1(c):
 Case III, with a gaussian form of evolution
is considered here for $\Omega=1, 0.1$ and $1+z_f=
1+z_s=10, 8$. The contours here are drawn in $\Omega_{IGM}--w$ space

\medskip
Figure 2: Temperature and Log$_{10}$ (${dT \over dz}$) for various heating
and cooling processes are plotted as functions of the redshift $z$ for
the Case I, a single power law luminosity evolution with $m=3.,
\Omega=0.1, \Omega_{IGM}=0.1, h=0.5, \beta_l=10^{-1.3}$. The solid curve
is Log$_{10}$ $T$; the dash and dot curve is due to cosmic ray heating;
the dotted, short dashed, long dashed curves denote cooling due
to recombination, expansion of the universe and line cooling
respectively.

\medskip
Figure 3: Fraction of the cosmic ray energy density that is lost
in moving through the IGM is plotted against the IGM density. The dotted
and dashed curves are for $\beta_l=10^{-1.5}, 10^{-1.3}$ respectively
for the case of a single power law evolution with $m=3, \Omega=0.1,
h=0.5$. With an intial lower cutoff of $30$ MeV ($\beta=10^{-0.6}$), these
curves then represent adiabatic loss of energy in galactic winds by
factors of $7.7, 5.0$ respectively.

\vfill\eject
\end